\documentclass[final,authoryear,5p,times,twocolumn]{elsarticle}

\usepackage{hyperref}

\usepackage{graphicx}
\usepackage{amssymb}

\journal{Astronomy and Computing}

\begin{document}

\begin{frontmatter}

\title{Probing the sparse tails of redshift distributions with Voronoi tessellations}

\author[brera]{Benjamin R. Granett}

\address[brera]{INAF Osservatorio Astronomico di Brera, via E. Bianchi 46, Merate Italy}
\ead{ben.granett@brera.inaf.it}

\begin{abstract}
We introduce an algorithm to estimate the redshift distribution of a sample of galaxies selected photometrically given a subsample with measured spectroscopic redshifts.   The approach uses a non-parametric Voronoi tessellation density estimator to interpolate the galaxy distribution in the redshift and photometric color space.    We test the  method on a mock dataset with a known color-redshift distribution.  We find that the Voronoi tessellation estimator performs well at reconstructing the tails of the redshift distribution of individual galaxies and gives unbiased estimates of the first and second moments.   The source code is publicly available at \url{http://bitbucket.org/bengranett/tailz}.
\end{abstract}

\begin{keyword}
galaxy surveys\sep statistical methods \sep photometric redshifts   \sep Voronoi tessellation density estimator
%% keywords here, in the form: keyword \sep keyword

%% MSC codes here, in the form: \MSC code \sep code
%% or \MSC[2008] code \sep code (2000 is the default)

\end{keyword}

\end{frontmatter}

% \linenumbers

\section{Introduction}

The photometric colors of a galaxy depend on its properties such as spectral energy distribution and redshift, but when analyzing photometric samples we are faced with the inverse problem: determine the redshift given only the observed colors.  This is an ill-posed problem since in general there is a distribution of redshift that can result in consistent color measurements; however, it is this redshift distribution that is the key to unlock the statistical power of photometric surveys \citep{Newman15}.  The accurate estimation of redshift distributions is an essential step of the analysis of photometric surveys and is necessary to extract cosmological measurements of the baryon acoustic scale, the growth rate of structure through redshift-space distortions as well as the signal encoded in the lensed shape correlations \citep{Mandelbaum08,Myers09,Wittman09,Asorey16}.  

Given a point in photometric color space, the redshift may be constrained by fitting against a library of spectral energy distribution templates \citep{Bolzonella00}.  This alone is not sufficient to extract the redshift distribution because we also need to know the relative abundances of the different galaxy types as a function of redshift \citep{Benitez00}.  For this reason, when representative samples exist, it is useful to empirically constrain the color-redshift distribution \citep{Wolf09}.

Empirical photometric redshift estimators begin with a sample of galaxies (the training set) of known redshift and photometric parameters such as color.  This sample is used to determine a mapping from the photometric parameter space to redshift.  We may imagine galaxies in a parameter space of color tagged with their respective redshifts.  Intuitively, given a new galaxy measurement with unknown redshift we may assume that it has similar properties to its neighbors in color space.  Then, its redshift can be predicted by selecting the set of nearest neighbors from the training set and measuring their mean redshift.  This idea underlies the nearest-neighbor algorithms adopted for photometric redshift estimation in the literature \citep{Csabai03,Lima08,Sheldon12,Geach12}, but other machine learning methods including  kernel regression and decision tree algorithms are also sensitive to the nearest neighbors \citep{Wang07,Gerdes10}.  These algorithms produce a single redshift estimate, but they may be adapted to estimate redshift distributions.  \citet{Ball08} perturb galaxies within their photometric error and measure the dispersion of the nearest-neighbor redshift estimate over a number of resamplings.  A variant of this approach is used in the random forest algorithm implemented in the Trees for Photo-Z (TPZ) code that we make use of below \citep{Kind13,Kind14}.
%For example, a photometric sample may be matched in color space to its nearest neighbors in the spectroscopic catalog to estimate the redshift distribution \citep{Csabai03,Lima08,Sheldon12,Masters15}.  The adaptive nature of these algorithms is important as galaxy colors can change abruptly with redshift due to the presence of strong spectral features.  These methods have been advanced with machine learning tools including kernel regression, prediction trees and neural networks \citep{Gerdes10,Collister04}.  Many estimators return only a single redshift estimate, but the redshift distribution may be probed by .     
%The spatial distribution of galaxies also encodes key information about the redshift distribution \citep{Schneider06,AragonCalvo15} but we will limit the  discussion to only photometric properties.  

\begin{figure}
\begin{center}
\includegraphics[width=3.in]{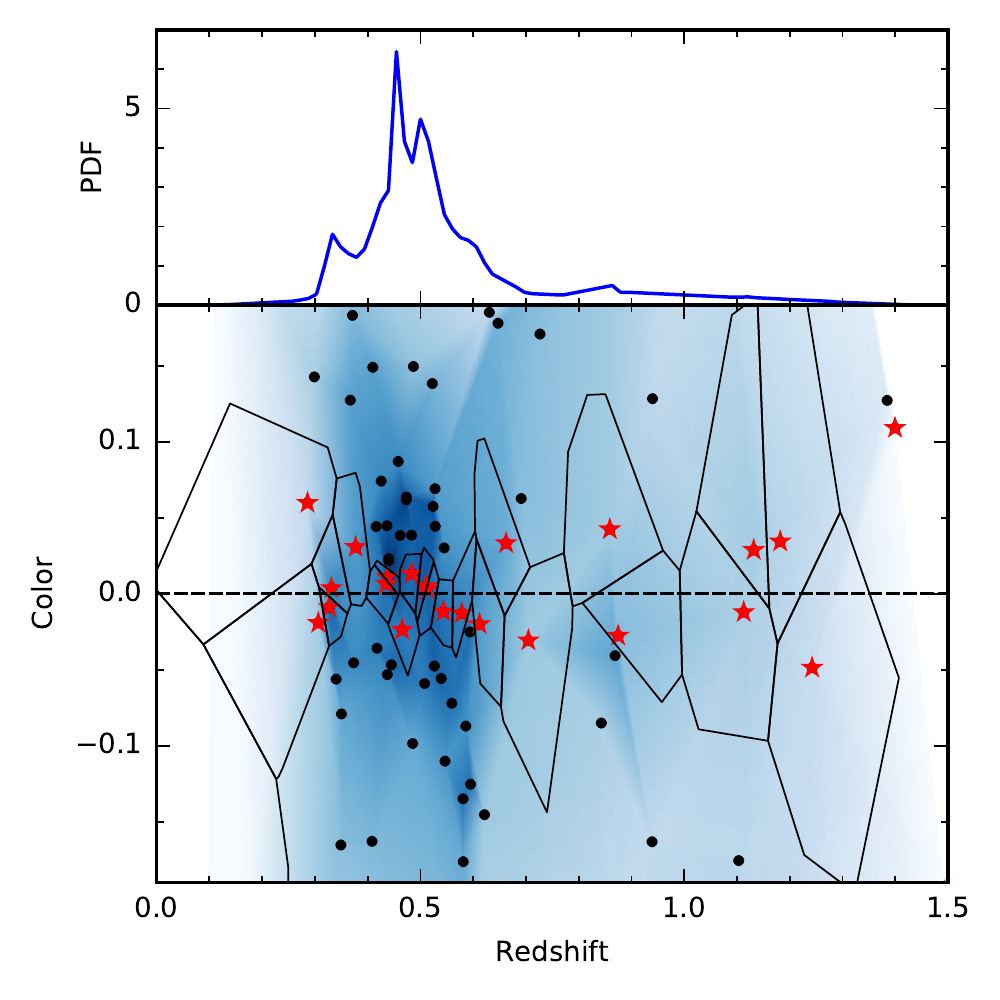}
\end{center}
\caption{The redshift probability distribution function (PDF) may be estimated empirically by interpolating the galaxy density in the space of color and redshift.  We illustrate a two-dimensional case and estimate the redshift  distribution along a line of constant color (dashed line).  The density is estimated from the inverse-volumes of Voronoi cells and is linearly interpolated to give a continuous  field (color gradient).  The resulting redshift distribution is shown in the top panel.  The galaxies used for interpolation are marked with stars and do not correspond to the nearest neighbors based on color distance.
  \label{fig:skewer}}
\end{figure}

Instead of considering redshift as a tag attached to each galaxy and building a mapping from color to redshift, we may imagine a continuous  density function in the parameter space of redshift and color.  Now the problem of estimating the redshift distribution becomes one of estimating this high-dimensional distribution function and then `skewering' it to measure the conditional density as a function of redshift at a fixed color.  This may seem daunting since galaxy surveys provide only a very sparse sampling of the parameter space, however we show that it may be solved by employing a Voronoi tessellation density estimator.  This follows from the common use of Voronoi tessellations for non-parametric density estimation \citep{Schaap00}.      The tessellation is constructed from a set of discrete points as illustrated in Fig. \ref{fig:skewer}.  A density may be assigned to each point computed from  the inverse of the volume of its associated Voronoi cell.  With this construction we  compute the density  as a function of redshift by interpolating the density field along a skewer.   In this way we probe even the very sparse tails of the redshift distribution.  In contrast, conventional nearest neighbor algorithms primarily sample from the peak of the distribution where the majority of neighbors lies.  Here, the points that influence the density estimator are distributed over the entire redshift range, as seen in Fig. \ref{fig:skewer}.

How may we assess the accuracy of an estimated redshift distribution?  One method that has been suggested is to check that the spectroscopic redshifts are consistent with the estimated redshift confidence intervals for a statistical sample \citep{Wittman16}.  However, with sparse training sets it can be difficult to adequately sample the tails of the distribution.  An alternative approach that we adopt is to test the recovery of a given known distribution. We construct a mock galaxy catalog by drawing redshifts and colors from a  distribution built from a Gaussian mixture model.  Then we quantify the accuracy of the estimator by comparing the first and second moments with the true values of the underlying distribution.  Our mock data mimics  the Canada France Hawaii Telescope Legacy Survey \citep[CFHTLS;][]{cfhtls} photometric sample and the VIMOS Public Extragalactic Redshift Survey \citep[VIPERS;][]{Guzzo14} spectroscopic subsample at redshift $0.5<z<1.2$.

\begin{figure}
\begin{center}
\includegraphics[width=3.5in]{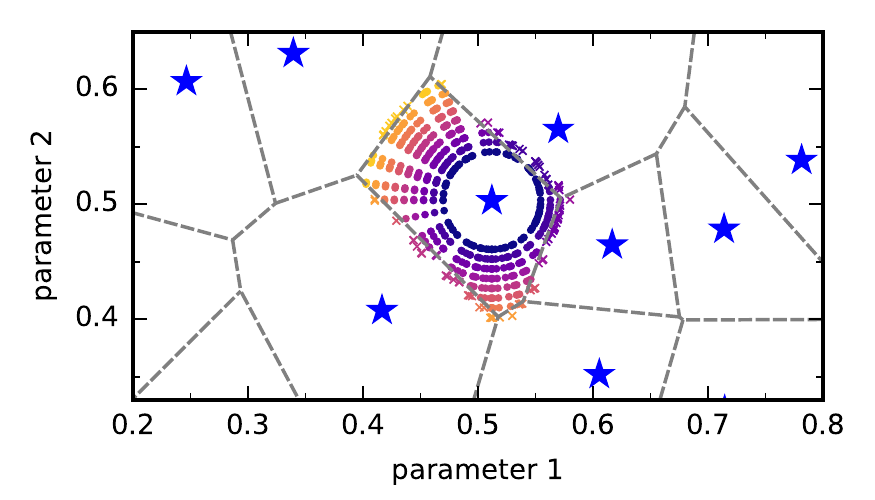}
\end{center}
\caption{Illustration of the Monte Carlo procedure to compute the volume of a Voronoi cell and identify its Delaunay neighbors. The galaxies used to generate the Voronoi diagram are marked with stars.  The volume of the Voronoi cell associated with a given galaxy is determined by sampling the cell with concentric spheres of test points.  The radius is increased iteratively and points that cross the boundary determined by the nearest neighbor are removed (x markers).  \label{fig:illus}}
\end{figure}

\section{Voronoi tessellation density estimator}
The Voronoi diagram partitions the $d$-dimensional space of galaxy properties into cells where each cell contains the region of the space that is nearer to the given galaxy than to any other.  The inverse of the cell volume gives an estimate of the local number density: $n_i = 1/V_i$.  We will use this density measure to estimate the parameter distribution function of galaxies.  

As the dimension of the point set increases, the number of Delaunay neighbors that surround each point grows.   In fact, the number of simplices in a Delaunay tessellation formed from $N$ points in dimension $d$ is of order $N^{d/2}$.  This makes it computationally intensive to not only construct tessellations in higher dimensions but also to store the results on disk.  We circumvent these problems by using  a Monte Carlo algorithm  to  estimate approximate volumes of the Voronoi cells by random sampling without the need to first compute the  tessellation.  This can be done because it is simple to determine if a point falls in a Voronoi cell by using a nearest-neighbor search without explicitly defining the cell boundaries.   We also use the procedure to determine the set of Delaunay neighbors which we may use to interpolate the density field.  The algorithm is not limited by the dimension, may be tuned to reach a desired precision and may be divided into a number of parallel tasks.  The steps are as follows.
\begin{enumerate}[Step 1.]
\item Select galaxy $i$ that will be the site of a Voronoi cell
\item Draw uniformly a sample of $N$ points on a hyper-sphere centered on the galaxy with radius $R_0$ set by half the distance to its nearest neighbor. 
\item Increment the radius $R_1=R_0+\Delta_R$ and move each test point to the larger radius.  
\item Find the nearest neighboring galaxy to each test point.  If a test point has crossed the boundary of the cell, its nearest galaxy will be a Delaunay neighbor.  Record the Delaunay neighbors and discard points that have crossed the cell boundary.
\item If test points remain, return to step 3.
\end{enumerate}

The construction of concentric spheres of test points is illustrated in Fig. \ref{fig:illus}.
The volume of the Voronoi cell may be estimated by making a weighted sum of all test points that were generated inside the cell.  In $d$-dimensions the volume of the hyper-ball constructed in step 1 is $V_{ball}(R_0) = a R_0^d$, where $a$ is a coefficient that depends on the dimension.  The volumes of the concentric shells on step $j$ are $V_{shell}(R_j) =  d a R_j^{d-1} \Delta_R$.  The total volume may then be estimated by
$V = V_{ball} + \sum_j \alpha_j V_{shell}(R_j)$ where $\alpha_j = m_j/N$ is the fraction of test points on the shell that remain inside the cell on iteration $j$.  The precision of the estimate will depend on the number of test points $N$ as well as the step size $\Delta_R$.

We impose a boundary so that exterior Voronoi cells are closed.  This is implemented by further removing test points on each iteration that fall outside the boundary.  These cells are flagged so that they may be given zero weight in the density analysis, but they still contribute to the analysis by bounding the volumes of galaxies that are interior.

The final density field estimator is given by  linear interpolation between galaxies.   This is carried out by first computing the $d$-dimensional Delaunay triangulation and interpolating the density between vertices \citep{numericalrecipes}.

\section{Recovery of a known distribution}
\subsection{Construction of mock data}

\begin{figure}
\begin{center}
\includegraphics[width=3.5in]{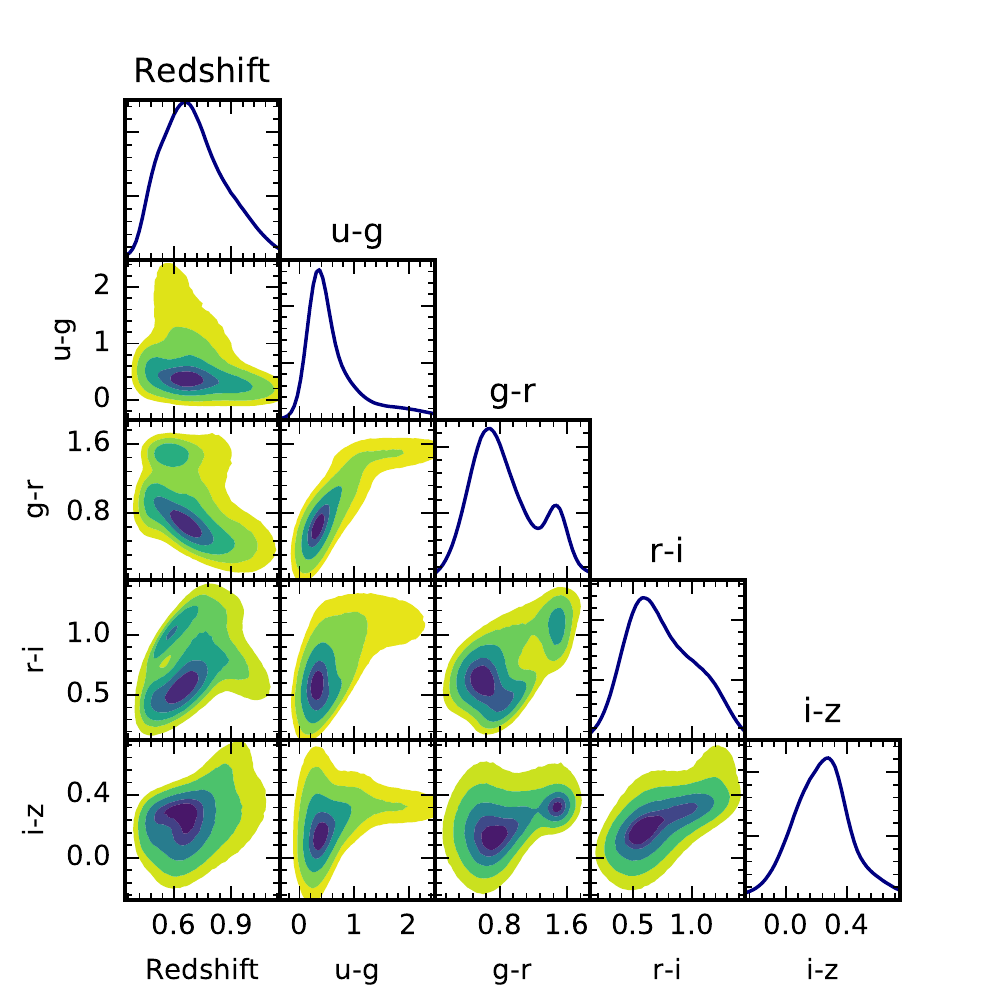}
\end{center}
\caption{The mock color-redshift distribution constructed from a Gaussian mixture model with eight components.  The model was fit to actual photometric and spectroscopic data from CFHTLS and VIPERS.  The shaded contours give the 90\%, 75\%, 50\%, 25\% and 10\% intervals.
 \label{fig:syn}}
\end{figure}

We construct a mock galaxy dataset by drawing samples from a well-defined color-redshift distribution.   For simplicity we use a Gaussian mixture model that may be expressed with an analytic function \citep{numericalrecipes}.

The model is fixed with photometric and spectroscopic data from CFHTLS and the VIPERS PDR-1 sample.  Four photometric colors are available including $u-g$, $g-r$, $r-i$ and $i-z$ as well as redshift for 45,476 galaxies.  The parameter space consists of five dimensions, and a point may be specified by a vector consisting of one redshift value and four colors: $\vec{x}=(z,c_1,c_2,c_3,c_4)$. We build up the distribution from a sum of Gaussian components with given mean $\vec{\mu}_k$, covariance $\Sigma_k$ and weight $p_k$:
\begin{equation}
f(\vec{x}) = \sum_{k=1}^{K} \frac{p_k}{\sqrt{\left(2\pi\right)^{d} \det{\Sigma_k}}} \exp{\left(-\frac{1}{2} (\vec{x}-\vec{\mu}_k)\Sigma_k^{-1}(\vec{x}-\vec{\mu}_k)^T\right)}.
\end{equation}
We use $K=8$ Gaussian components.  This choice provides a reasonable fit by eye and also corresponds to the point where the Akaike information criterion begins to flatten.  The fitting is carried out using  the Gaussian mixture model implementation in the  Scikit-learn  library for Python\footnote{\url{http://scikit-learn.org}} \citep{scikit-learn}.  

A mock galaxy may be drawn from this distribution by first selecting one of the $K$ components with probability $p_k$ and then drawing a set of values $\vec{x}$ from the specified multivariate Gaussian distribution.  The derived mock distribution is illustrated in Fig. \ref{fig:syn}.  We sample the distribution to generate a mock catalog with 50,000 galaxies and an independent test sample of 1,000 galaxies.

\subsection{VT implementation}
Following the Monte Carlo algorithm described above, we sample each Voronoi cell using $N = 10^4$ uniformly distributed test points on a hyper-sphere.  The sphere begins with radius $R_0$ set by half the distance to the nearest neighbor.  The radius is then incremented by $\Delta_R=R_0/10$ on each iteration.  The normalized redshift distributions are computed over the range $0<z<2$ with interval $\Delta_z=0.001$ and then resampled to $\Delta_z=0.01$.

\subsection{TPZ code}
For the purpose of comparison we run the Trees for Photo-Z (TPZ) code version 1.2 \citep{Kind13,Kind14b} on the mock dataset.  This code implements a prediction tree algorithm and also constructs individual galaxy redshift distributions by using the random forest technique.  We set a constant color measurement error of $\sigma_c=0.03$ magnitudes and in total we realize 5,000 trees ({\tt NRandom}=10; {\tt NTrees}=500).  

\subsection{Results}

\begin{figure*}
\begin{center}
\includegraphics[scale=0.9]{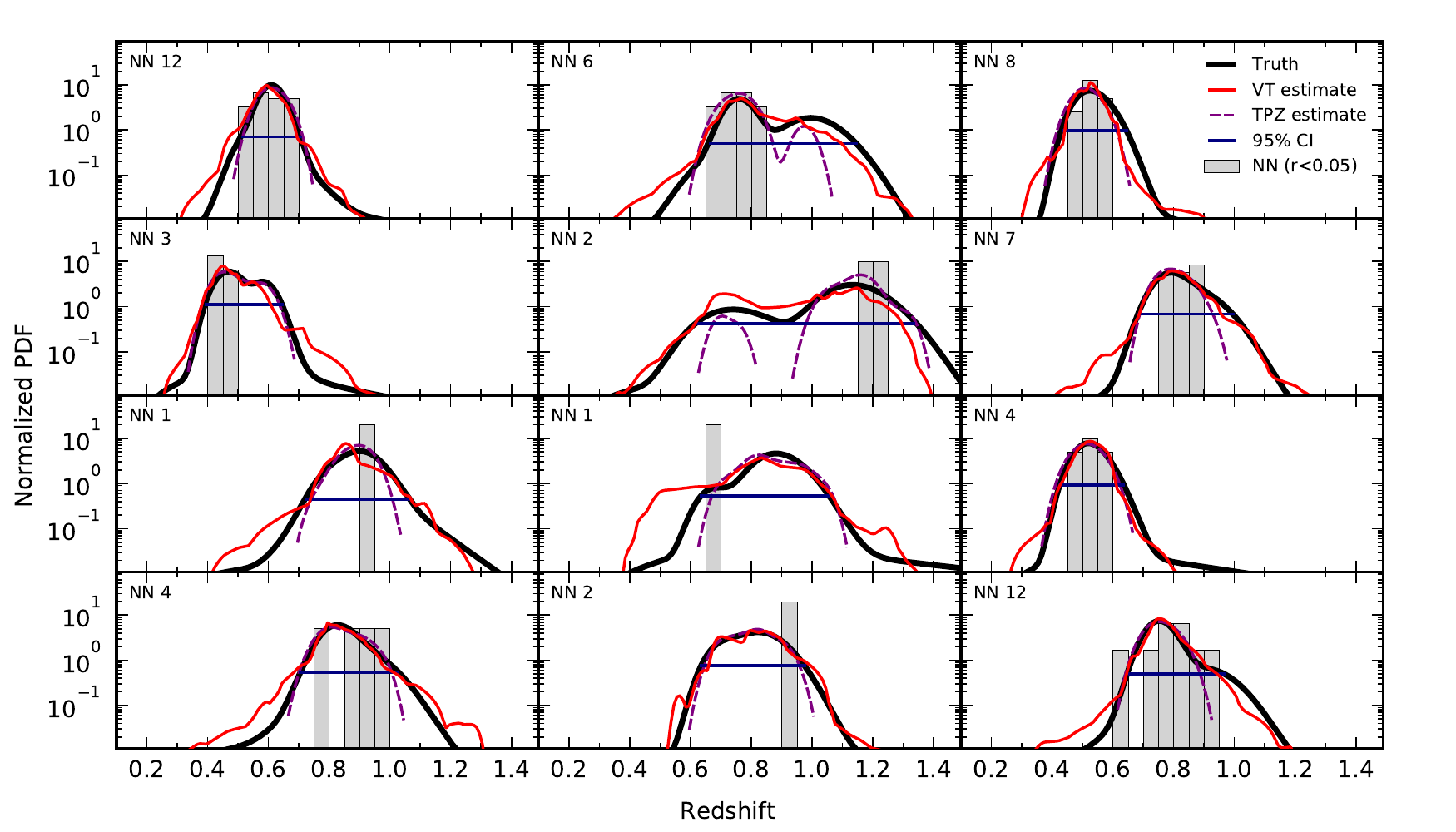}
\end{center}
\caption{ Mock redshift distribution estimates for 12 galaxies selected at random from the test sample.  The thick black curve gives the true redshift distribution known analytically and the 95\% confidence interval is indicated by the horizontal line.  The shaded histograms show the distribution of redshifts for the sparse sample of nearest-neighbor galaxies within a radius of 0.05 magnitudes in the color space.  The thin red curve shows the Voronoi tessellation (VT) estimate of the redshift distribution.  The dashed curve shows the estimate given by the Trees for Photo-Z (TPZ) code.  \label{fig:examples}}
\end{figure*}

We estimate the redshift distributions for the testing set of 1000 galaxies and compare the two methods: Voronoi tessellation (VT) and Trees for Photo-Z (TPZ).
Fig. \ref{fig:examples} shows the results from 12 galaxies selected at random.     We find good agreement between the Voronoi tessellation estimate and the true distribution, despite the sparsity of the sample.  The distributions predicted by  the TPZ algorithm also match the peaks well, but in some cases they do not show support over the full 95\% confidence interval.  

We compute the first and second moments of the distribution to quantify the performance of the estimators.  These correspond to the mean and standard deviation:
$\bar{z} = \sum_i z_i p_i$,  $\sigma_z = \sqrt{\sum_i (z_i-\bar{z})^2 p_i}$.

\begin{figure}
\begin{center}
\includegraphics[width=3.5in]{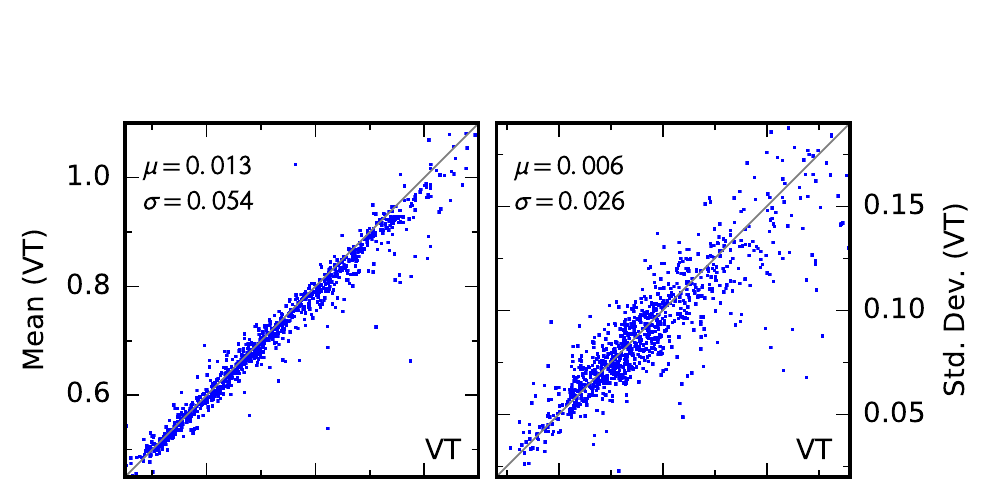}
\includegraphics[width=3.5in]{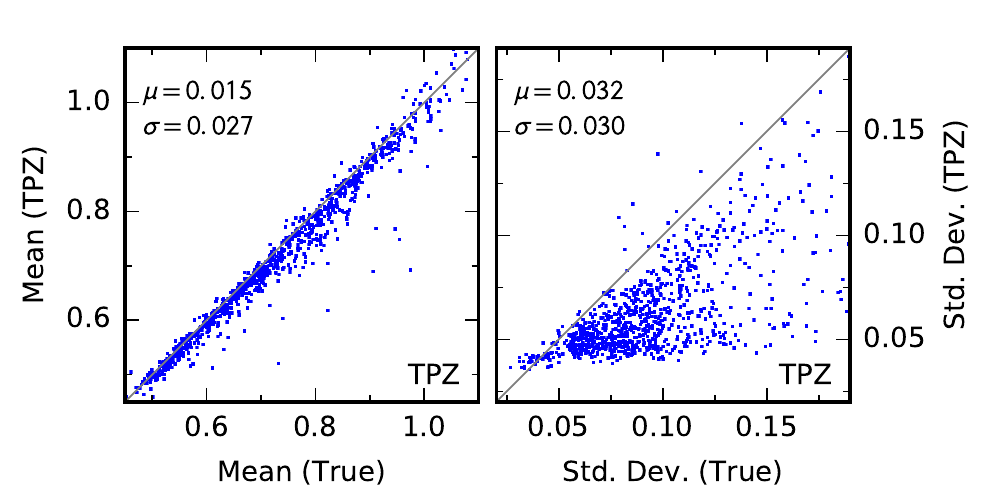}
\end{center}
\caption{The first and second moments of the redshift distributions estimated with the Voronoi tessellation (VT, top) and Trees for Photo-Z (TPZ, bottom) algorithms.  \label{fig:moments}}
\end{figure}

In Fig. \ref{fig:moments} we compare the estimates of the moments against the true values known for the distributions.   Both estimators give unbiased estimates of the mean redshift, and we find that the TPZ  code gives lower dispersion.  However, the second moment is systematically underestimated by the TPZ code while it is recovered without bias by the Voronoi tessellation estimator.

\section{Discussion and Conclusion}
Empirical photometric redshift estimators typically work by identifying nearest neighbors within the photometric parameter space.  These neighbors are likely to inhabit the peak of the redshift distribution where the density of galaxies is highest.  This leads to underestimation of the tails.  We have presented an algorithm which aims to directly estimate the full redshift distribution based upon a Voronoi tessellation density estimator.  The Voronoi method was selected because it has no free parameters, but other density estimators may be implemented as well.  We test the performance of the method on a mock dataset for which the redshift distribution as a function of galaxy color is known analytically.   We demonstrate that the first and second moments of the distribution are estimated without bias.  

The test distribution we used is idealized in that it is made up of smooth Gaussian components.  In reality the color-redshift distribution of galaxies is not Gaussian and  shows sharp features where spectral lines enter a photometric band \citep{Masters15}.  Additionally, samples dominated by cosmic variance can display strong clustering as a function of redshift.  However, in wide-area datasets, and considering photometric uncertainties, the smooth approximation is reasonable.  Nevertheless, it will be important to assess the performance of the redshift distribution estimates on actual spectroscopic survey data. 

The Voronoi tessellation algorithm can be readily adapted to spectroscopic survey data. We have neglected the effects of photometric errors and effectively assumed they are constant.  In reality, photometric uncertainties increase at fainter flux and we can expect that the redshift distributions will broaden.  The effect may be accounted for by changing the parameter space of our estimator.  Instead of computing density in the redshift-color space, we may use the distribution of redshift and magnitude.  In general we may incorporate any photometric parameter that is correlated with redshift.   

The survey selection function raises further challenges for empirical methods.   It is often the case that there are systematic differences in selection between spectroscopic and photometric samples \citep[see VIPERS;][]{Guzzo14}.  Using the Voronoi tessellation estimator we may account for these biases by appropriately weighting  galaxies in the training set  such that the distributions of photometric properties match.  This approach to correcting redshift distributions was studied by \citet{Lima08}.  

We have applied the Voronoi tessellation estimator to a dataset with galaxy redshift and  four  colors; given the dimensionality of photometric data, will it be feasible to apply in higher dimensions?  The `curse of dimensionality' enters when computing the volumes of Voronoi cells.  As the dimension is increased, cells become larger, the number of neighbors grows and we expect a larger fraction of cells will be unbounded.  A solution is to reduce the dimensionality of the problem using techniques such as principal component analysis and self-organizing maps.   This is feasible since photometric parameters generally contain redundant information.  A second option is to modify the algorithm with fast heuristics that provide approximate results.  One approach may be to substitute the Voronoi volume with the volume of the hyper-ball that extends to the $N^{\rm th}$ nearest neighbor.

Upcoming imaging surveys will be faced with the challenge of constraining redshift distributions of photometric samples with limited spectroscopic measurements.  We can benefit from methods that exploit the full parameter space of galaxy redshift and photometric properties in particular to estimate the sparse tails of the distribution.  The code developed for this work is publicly available in the project repository:  \url{http://bitbucket.org/bengranett/tailz}.

 \section*{Acknowledgement}
I thank Angela Iovino and Adam Hawken for stimulating discussions about tessellations  and Jason Dossett for his support of the Dorami computer cluster.  The VIMOS Public Extragalactic Redshift Survey (VIPERS) PDR-1 sample was prepared by the VIPERS team.  VIPERS has been performed using the ESO Very Large Telescope, under the ``Large Programme'' 182.A-0886. The participating institutions and funding agencies are listed at \url{http://vipers.inaf.it}.  BRG receives financial support from the European Research Council through the Darklight ERC Advanced Research Grant (\# 291521).

\bibliographystyle{hapj}
\bibliography{photz}

\end{document}